\documentclass[aps,pre,onecolumn,superscriptaddress,showkeys,floatfix]{revtex4}
\usepackage[]{graphicx}
\usepackage[]{subfigure}

\usepackage{amsmath}%
\usepackage{amsfonts}%
\usepackage{amssymb}%
\begin{document}

\title{Search for an optimal light-extracting surface\\
derived from the morphology of a firefly lantern}

%>>>> The author is responsible for formatting the
%  author list and their institutions.  Use  \skiplinehalf
%  to separate author list from addresses and between each address.
%  The correspondence between each author and his/her address
%  can be indicated with a superscript in italics,
%  which is easily obtained with \supit{}.

\author{Annick Bay}\affiliation{Research Center in Physics of Matter and Radiation
(PMR),\\University of Namur (FUNDP), 61 rue de Bruxelles,
B-5000 Namur Belgium}\email{annick.bay@fundp.ac.be}
\author{Micha\"{e}l Sarrazin}\affiliation{Research Center in Physics of Matter and Radiation
(PMR),\\University of Namur (FUNDP), 61 rue de Bruxelles,
B-5000 Namur Belgium} %\email{michael.sarrazin@fundp.ac.be}
\author{Jean Pol Vigneron}\affiliation{Research Center in Physics of Matter and Radiation
(PMR),\\University of Namur (FUNDP), 61 rue de Bruxelles,
B-5000 Namur Belgium} %\email{jean-pol.vigneron@fundp.ac.be}

%%%%%%%%%%%%%%%%%%%%%%%%%%%%%%%%%%%%%%%%%%%%%%%%%%%%%%%%%%%%%
%>>>> uncomment following for page numbers
% \pagestyle{plain}
%>>>> uncomment following to start page numbering at 301
%\setcounter{page}{301}

%%%%%%%%%%%%%%%%%%%%%%%%%%%%%%%%%%%%%%%%%%%%%%%%%%%%%%%%%%%%%%%%%%%%%%%%%%%%%%%%%%%%%%%
\begin{abstract}
Fireflies lighten up our warm summer evenings. There is more physic
behind these little animals than anyone of us could imagine. In this
paper we analyze from a physical point of view one structure found
on the firefly lantern, the one which best improves light
extraction. Moreover, simulations will be done to show why this
specific structure may be more effective than a "human-thought" one.

\end{abstract}

%>>>> Include a list of keywords after the abstract


\maketitle

%\begin{figure}[t]
%\begin{center}
%\begin{tabular}{c}
%\includegraphics[width=10cm]{fig1.eps}
%\end{tabular}
%\end{center}
%\caption[example]
%>>>> use \label inside caption to get Fig. number with \ref{}
%{ \label{fig1} }
%\end{figure}
%--------------------------------------------------------------------------------------
%       Introduction
%--------------------------------------------------------------------------------------

\section{Light in Nature}

Light is a major path of commuication in nature. Since the emergence
of the eye in the animal kingdom, the variety of species has found
new opportunities to grow. Parker \cite{Parker-book-2003} suggested
the "light switch argument" in order to explain this known "Big Bang
of the evolution". With vision, animals -- especially insects -- had
either to adapt their outer appearance to be seen or hide effectively.
For intra-specific communication, males have to wear bright colors
that suggest genetic richness to attract the females. In contrary,
for inter-specific communication, insects often have to make
themselves invisible to escape predators. Another way to discourage
predators is to show bright signal colors, such as red or yellow,
which is commonly known to signal toxicity in nature. All these ways
of communication use sunlight. The importance of photonic
communication between animals and their environment is best
underlined by what happens when sunlight disappears, at night or in
the deep sea. Then and there the communication is still based on
vision and animal produce their own light. Deep-sea fishes use
mainly blue light to attract their preys and on earth, when the sun
sets, insects starts to lighten up. Different species of tropical
click-beetle show three light-emitting spots on their back;
fireflies emit light at all stages of their evolution. The eggs and
the larvaes lighten up to prevent a predator attack and as adults
use bioluminescence to improve mating occurrences.

Fireflies produce their own light by a catalyzed oxidation chemical
reaction, which is called bioluminescence. This bioluminescence
reaction is well known \cite{Vico-jpp-2008} : it is even possible to
recreate it in vitro \cite{Alam-NL-2012} . In view of its
importance, the structure of the bioluminescent organ of the firefly
has received the considerable attention
\cite{Ghiradella-I-1998,Hanna-JUR-1976,Locke-JCB-1971,Smalley-JHC-1980} .
The present paper is based on the results from a former article by
Bay and Vigneron \cite{Bay-SPIE-2009} . There will be a brief recall
of the results. At first we give a brief account of the main results
in this earlier work, which then is augmented by new data that might
be useful to technical and engineering developments.

%--------------------------------------------------------------------------------------
%       Morphology former results
%--------------------------------------------------------------------------------------
\section{Firefly morphology and its influence on light extraction}

Morphological analysis by scanning electronic microscopy (SEM) has
revealed different structures in and around the bioluminescent organ
of the fireflies. Two of these structures seemed to be particulary
interesting: (1) the spheres in the bioluminescent organ and (2) the
tilted scales on the outer cuticle above the bioluminescent organ. A
detailed model, that combined these two structures led to a improved
light extraction which reaches a factor two compared to a flat
surface reference. We will here analyse more precisely the impact of
the tilted scales geometry on light extraction.

The tilted scales geometry has two distinct effects on the light
extraction: (1) The interface is not plane anymore, but tilted which
could help changing the critical angle phenomenon. (2) The
protruding ends create sharp edges on which diffusion can take
place. From the SEM analysis an average height of the protruding end
of 3 $\mu$m and a periodicity of 10 $\mu$m could be determined.
These mechanisms have received an appropriate modeling in all
details in two dimensions \cite{Bay-SPIE-2009} : the vertical corrugation profile (z
direction) in this model varies only in the lateral
(y direction), while the height stays invariant in the x direction.

\begin{figure}[h]
\centering \subfigure[Outer cuticle of the firefly showing the
misfitting scales.]{\includegraphics[height=4.5cm]{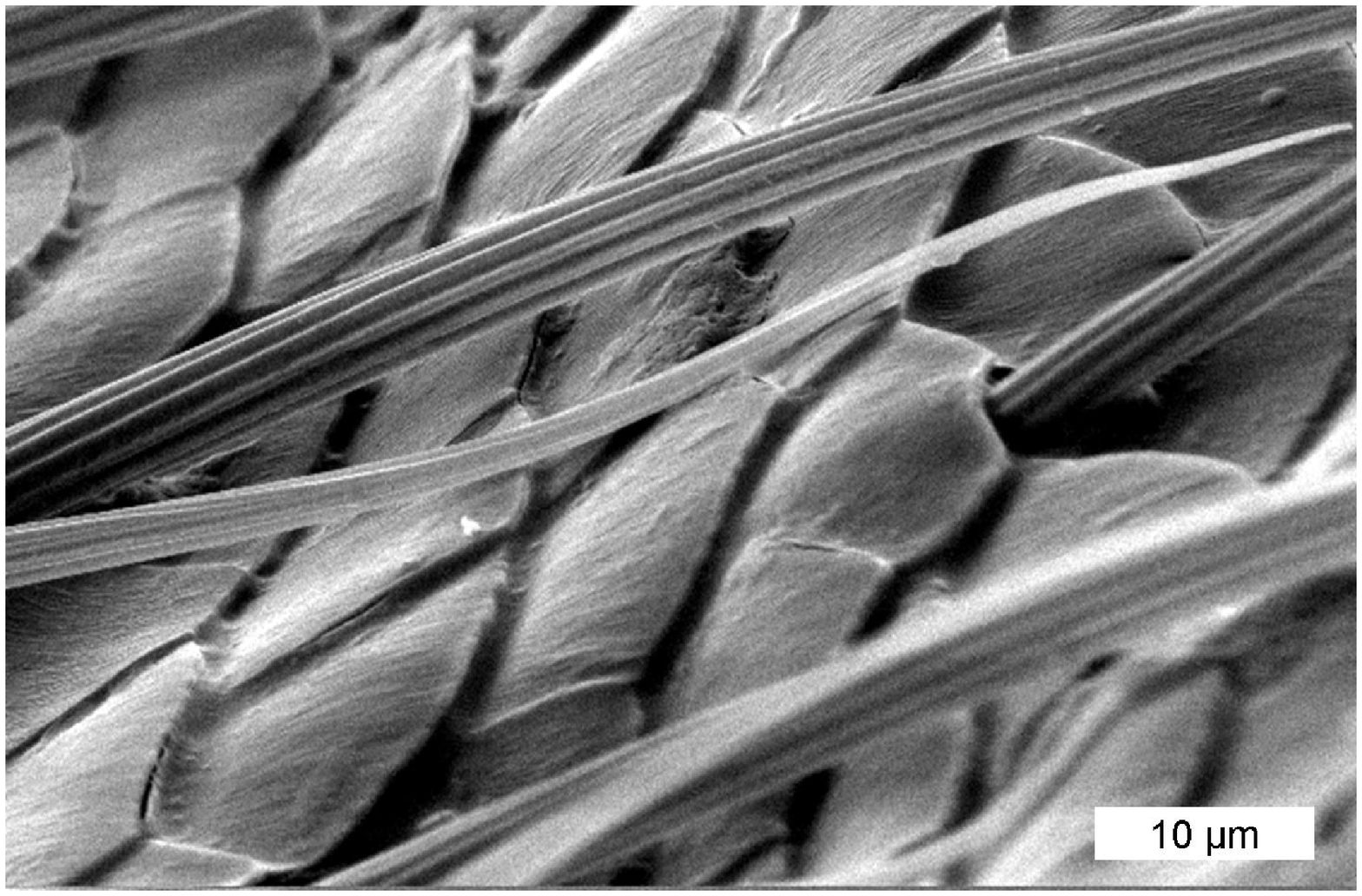}}\qquad
\subfigure[Grating with a triangular profile, with period of $10~\mu$m
and height of $3\mu$m, simulating the scales with a border misfit.]
{\includegraphics[height=4.5cm]{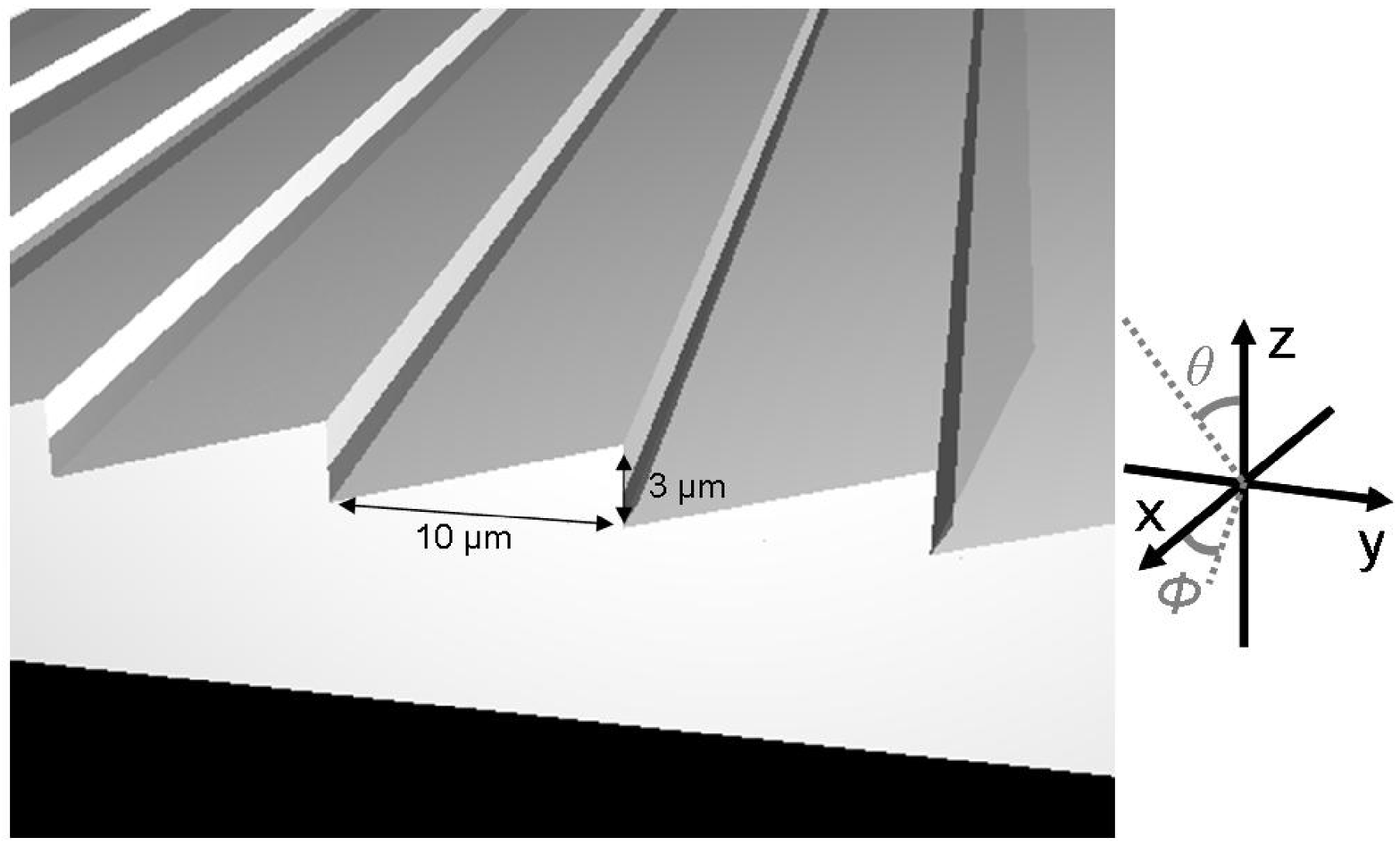}}\\
\caption{Structure found on the outer surface of the firefly-cuticle}
\label{fig6}
\end{figure}

\subsection{Light extraction path}

\begin{figure}[b]
\begin{center}
\includegraphics[height=8cm]{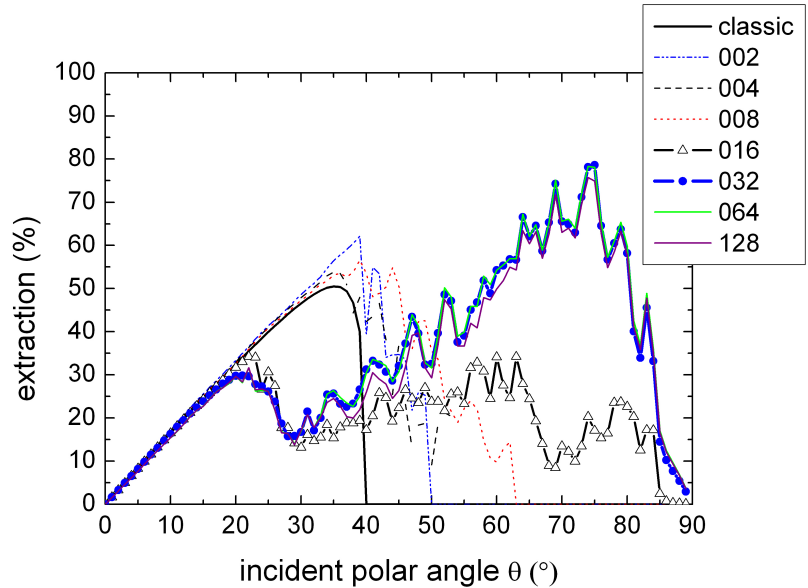}
\caption{Convergence study of the tilted scales. The incident
azimuthal angle has been fixed perpendicular to the tilted edges ($\phi=90^\circ$).
The incident polar angle varies from $0^{\circ}$ to $90^{\circ}$. 32
plane waves are enough to show properly the effect of the tilted
scales. The critical angle is lost and a huge part of the light is extracted at high incident polar angles.} \label{fig_convergence}
\end{center}
\end{figure}

By studying the convergence of the light extraction as a function of
the accounted number of diffraction orders in this specific
structure, some precious information about the way light escapes can
be obtained. New simulations have been performed with the
three-dimensional transfer matrix algorithm \cite{Vigneron-SPIE-2006} .
 The incident azimuthal direction has been
fixed perpendicular to the protruding edges ($\phi = 90^\circ$) and
the incident polar angles ($\theta$) varies from $0^{\circ}$ to
$90^{\circ}$. This simulation is conducted with a wavelength of 560
nm, close to the emission peak of the firefly \cite{Hastings-Gene-1996} . The incident light is considered
depolarized, which is achieved by averaging TE and TM polarizations.
With a period (10 $\mu$m) much longer than the wavelength,
transmission can occur via many diffraction orders with emergences
widely distributed in space. If in the calculation we artificially
limit the number of diffraction orders accounted for, we can examine
in detail the paths of the emerging light. Fig. \ref{fig_convergence} shows that for a limited number of diffraction
orders (2 - 4 emerging plane waves), the tilted scales are
hardly considered by our computation and the critical angle is
similar to the one in a planar interface (classic). With more
diffraction orders (8 - 16), the critical angle increases and
disappears even completely (32-128). Low diffraction orders cannot
take into account the influence of the sharp edge at the protruding
end of the scales. With more diffraction orders the effect of the
prism can be clearly seen. Remarkably, a huge part of the light is
extracted at high angles. This light is completely lost in the
plane-interface case. The convergence study shows as well, that 32
plane waves can already properly account for the effect of the
corrugation.

\subsection{Light extraction map}

\begin{figure}[h]
\begin{center}
\includegraphics[height=6.5cm]{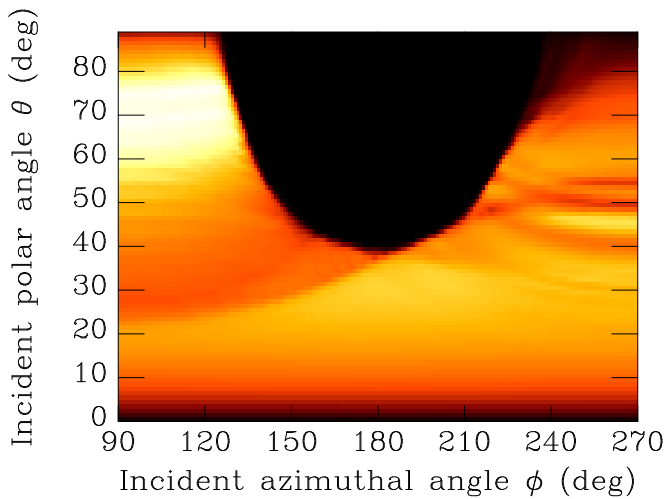}
\end{center}
\caption{Map of the relative extraction intensity in function of the
incident polar angle and the incident azimuthal angle. The light is
highly extracted when the lightrays are impinging on the surface
perpendicular to the edge of the protruding structure and
perpendicular to the slope. The color scale is relative: black
shows zero transmission, and white shows maximal transmission.}
\label{scales_map}
\end{figure}

For a full understanding of the effect of the sharp edge effect, we
calculate the light extraction for varying incident polar and
azimuthal angles. Polar angles ($\theta$) range from $0^{\circ}$ to
$90^{\circ}$ and azimuthal angles ($\phi$) from $90^{\circ}$ to $270^{\circ}$,
expandable to the full range ($360^\circ$) by symmetry. The
significance of these azimuthal angles are as follows: for
$\phi=90^\circ$ the light impinges perpendicular to the sharp edge,
for $\phi=180^{\circ}$ the light impinges parallel to this edge and
for $\phi=270^{\circ}$ the light impinges perpendicular to the
slope. The map in Fig. \ref{scales_map} shows the extracted light
intensity as a function of the incident polar and azimuthal angle.
Parallel to the edge ($\phi=180^{\circ}$), the total reflexion above
the critical angle reduces the extraction to zero. While getting
more and more perpendicular to the sharp edge ($\phi=90^\circ$ or
$\phi=270^\circ$), this effect is attenuated and the transmission is
highly enhanced: the total reflexion disappears completely. One can
see that the light extraction efficiency is not symmetric: light
escapes more easily the structure in the direction of the sharp edge
($90^{\circ}$) than in the direction of the slope ($270^{\circ}$).

This enhancement is achieved along two paths: (1) the slope of the
tilted scale changes completely the geometry of the surface and the
effect of the critical angle is therefore perturbated. (2) A strong
diffusion phenomenon takes places on the sharp edges of the tilted
scales. The size of this grating allows a lot of diffraction orders
to escape the source volume. We rather talk about diffusion instead
of diffraction.

\subsection{Search for the optimal geometry}

\begin{figure}[tbp]
\begin{center}
\includegraphics[height=9cm]{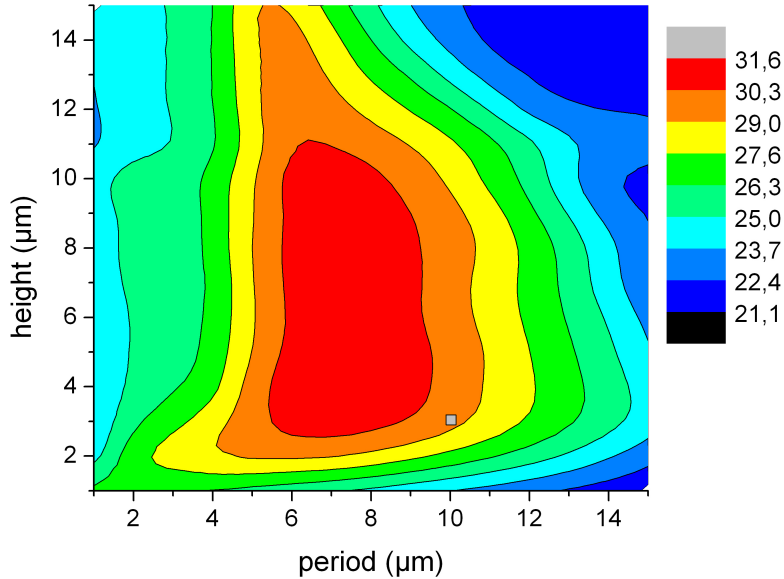}
\caption{Optimal geometry for the tilted scale (or the so-called firefly) geometry. The map shows the light extraction percentage as a function of the period and the heigth. The grey rectangle shows the geometry of the
firefly.}\label{opt_scales}
\end{center}
\end{figure}

Another interesting question concerns optimality: has the firefly
done "the best she could" with this structure to improve on the
light extraction. From the biomimetic point of view, it would be
interesting to know if another geometry could give even better results.
Therefore simulations with various values of the period and the
prism height were carried out with both parameters ranging from 1
$\mu$m to 15 $\mu$m. TE and TM polarizations have been computed, but in view
of the slight difference obtained on these maps, only the TE maps
are shown in Fig. \ref{opt_scales}. There is a slight difference
between the optimization calculation for the TM polarization, but
the difference is not significant to change the result completely.

The map on Fig. \ref{opt_scales} shows the extracted light intensity
as a function of the period (x-axis) and the corrugation height
(y-axis). One can see clearly that the firefly is not far from the
optimal case (grey rectangle on the map). In fact the differences
between the highest light extraction intensity and the light
extraction achieved by the firefly is less than 2\%. This small
discrepancy is remarkable and suggests that the optimization of the
corrugation in the firefly is dominated by the light extraction
process. This is astonishing, because (1) the cuticle not only has
to be easily permeable for light, but it has also to show different
properties, such as lightness, mechanical stiffness and
hydrophobicity. The cuticle is then not only optimized for light
extraction but more multi-optimized for a lot of different aspects.
(2) We used many different specimens at different stages of their
life in order to get average values of the geometrical parameters
and the variability is at least of the order of these 2\%.

The data contained in Fig. \ref{opt_scales} and their adaptation to
inorganic materials is of value for many light extraction designs
including artificial ones. The originality of this work lies in its
biomimetic approach which directly points to the size of the
structures and an optimal profile. Instead of figuring out new
designs from scratch, we use readily optimized solutions from
evolution and inspiration from nature. As other living
organisms \cite{Vigneron-PRE-2005,Parker-nat-2003,Vukusic-nat-2000,Zi-PRE-2006} ,
fireflies have benefitted from the trials and errors during the long
process of evolution in order to find an adapted structure to
improve light extraction. Engineered optimization only looks back on
several decades. It is difficult to tell whether such a structure
with their optimal parameters could have emerged without the natural
model. It seems, that earlier studies show solutions usually of size
smaller than the one revealed here. Considerations of
photonic-crystal films and other structures which show
inhomogeneities of the scale of the light wavelength are most often
worked out \cite{Jin-OE-2012, Lai-DRM-2011, Uthir.-MSEB-2010} .

The solution indicated by the firefly is actually multi-scale, as
the period chosen is much larger, than the emitted wavelength and
the abruptness of the edge responsible for the diffusion corresponds
to a length scale much smaller than the wavelength. Consistent with
this structural characteristics, other structures might be designed
and checked for efficiency.

%--------------------------------------------------------------------------------------
%       Other structures
%--------------------------------------------------------------------------------------

\section{Derived bioinspired structures}

Regarding these huge improvements of the prism-shaped form, we could
now think about similar structures, like sharp triangles, pyramids
or cones. Intuitively those structures could even be better for
improving light extraction. Such structures present sharp edges and
at the same time provide a gradual adaption of the refractive index.

\subsection{New morphologies}

\begin{figure}[tbp]
\centering
\includegraphics[height=9cm]{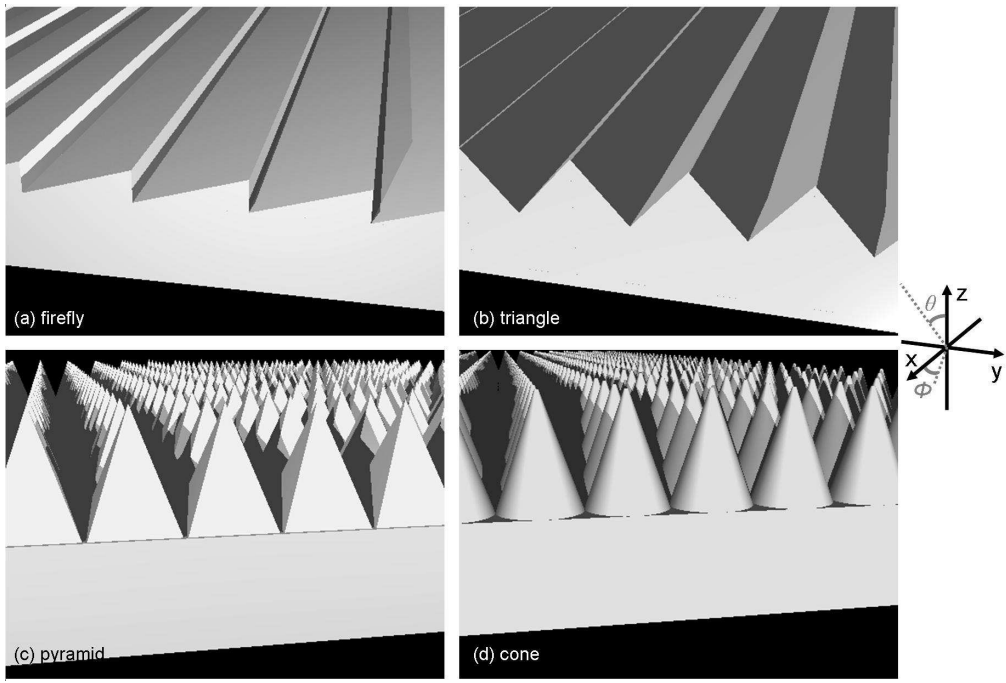}
\caption{New morphologies}
\label{structures}
\end{figure}

Fig. \ref{structures}(a) shows the structure found on the abdomen
of the firefly which has been studied earlier. After this, we will
consider three new specific structures:
\begin{enumerate}
\item Two-dimensional triangle-shaped structure. This structure is quite similar to the one found on the firefly, but shows a symmetry around the z-axis. Similar to the firefly, in the direction of the x-axis the corrugation exhibits a total translational invariance (Fig. \ref{structures}(b)).
\item Three-dimensional pyramid structure. This shape is similar to the two-dimensional triangle structure, but varies as well along the x-axis (Fig. \ref{structures}(c)).
\item Three-dimensional conical structure. This structure is similar to the three-dimensional pyramid structure, but due to the conical shape, no sharp corners appear and the variation of the refractive index is smoother (Fig. \ref{structures}(d)).
\end{enumerate}

We will proceed to the same calculations as before, i.e. check the
convergence of the calculations, define the optimal geometry and
show the map of the relative extraction in function of the incident
polar and azimuthal angle. For an easier comparison, the former
results will be recalled next to the new simulations.

\subsection{New light extraction paths}

\begin{figure}[tbp]
\begin{center}
\includegraphics[height=10cm]{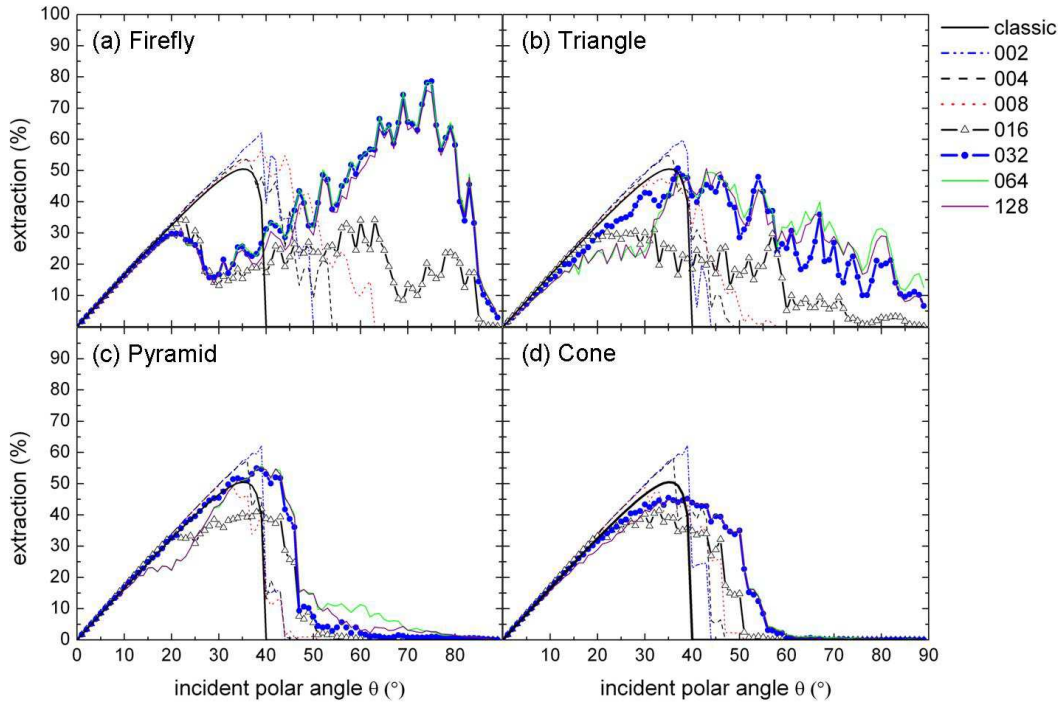}
\end{center}
\caption{Convergence calculation for the different structures. The incident
azimuthal angle has been fixed to $\phi=90^\circ$.
The incident polar angle varies from $0^{\circ}$ to $90^{\circ}$. 32
plane waves are enough to show properly the effect of the different structures.}
\label{convergence}
\end{figure}

A complete convergence study has been performed for these new
structures. We can keep with 32 plane waves to properly account for
the light propagation in the structure as justified in the following (Fig. \ref{convergence}). The simulations have been done, as before, at a wavelength of 560 nm and for depolarized
light. The azimuthal angle is taken to be at $90^\circ$ for all
calculations. We can already see by inspection of these graphs, that
the firefly has the most effective light extractor. The light can
easily escape the material beyond the critical angle (Fig.
\ref{convergence}(a)) and the light extraction is considerably
enhanced. The triangle structure wipes out the effect of the
critical angle as well (Fig. \ref{convergence}(b)), but the light
extraction is not improved with the same effectiveness as in the
firefly case. The pyramid and the conical structures also improves
on the extraction angle, but the intensity, in this specific
direction, is way less effective than the firefly structure.

%\subfigure[Firefly]{\includegraphics[height=4.5cm]{conv_scales.eps}}\qquad
%\subfigure[Triangle]{\includegraphics[height=4.5cm]{conv_tri-2D.eps}}\\
%\subfigure[Pyramid]{\includegraphics[height=4.5cm]{conv_tri-3D.eps}}\qquad
%\subfigure[Cone]{\includegraphics[height=4.5cm]{conv_d-sph.eps}}\\

\subsection{Search for the new optimal geometry}

\begin{figure}[tbp]
\centering
\includegraphics[height=10cm]{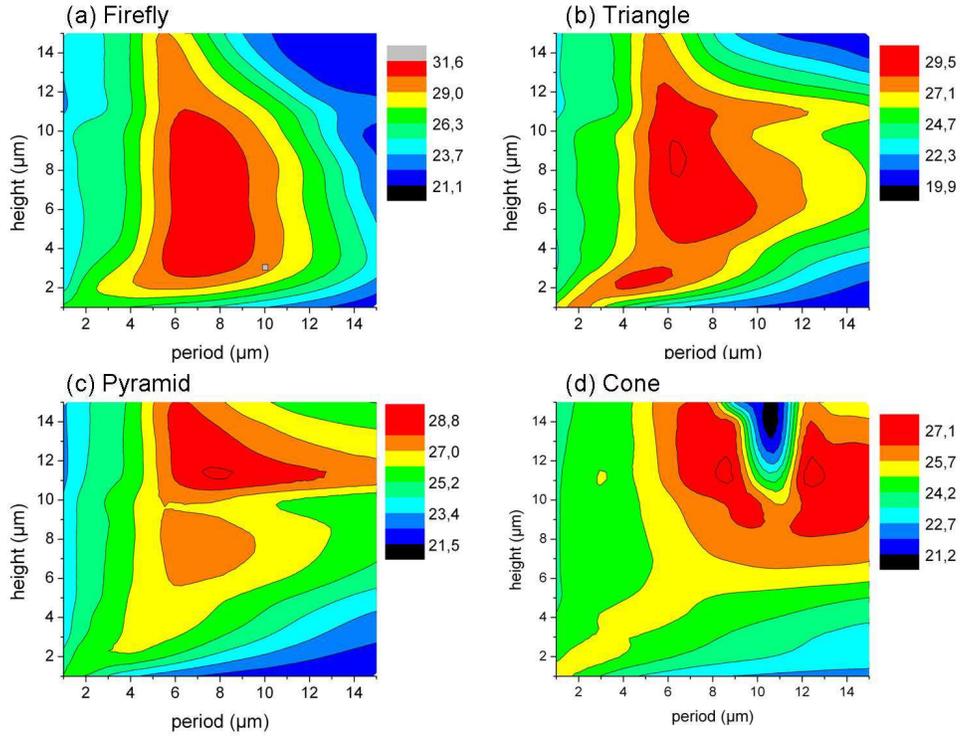}
\caption{Optimal geometry for the different structures. Maps showing the light extraction percentage as a function of the period and the heigth.}
\label{Optimisation}
\end{figure}

The maps on Fig. \ref{Optimisation} show the integrated extraction
coefficient in function of the periodicity (x-axis) and the height
(y-axis) of the geometry.

There is little difference between the firefly case (Fig.
\ref{Optimisation}(a)) and the two-dimensional triangle (Fig.
\ref{Optimisation}(b)). The firefly structure is more efficient over
a wider range of heights where the triangle structure needs bigger
heights for efficient extraction. However the highest value on the
scale shows only 29.5\% of light extraction for the triangle shape,
whereas the firefly-shape reaches 31.6\% at its maximum. The firefly
is still slightly better.

Fig. \ref{Optimisation}(c) shows that the highest extraction values
for the pyramids are located at larger heights than the previous
ones. Again, the optimal range is slightly smaller. Moreover the
light extraction reaches only 28.8\% in this case. The highest value occurs for a period of 8 $\mu$m and a
height of 11 $\mu$m. The map of the conical structures (Fig. \ref{Optimisation}(d)) is quite different to the pyramid structure. The extraction efficiencies are a little bit lower (highest values at 27.1 \%).
The highest extraction values occur at a period of 8 $\mu$m and a
height of 11 $\mu$m and at a period of 9 $\mu$m and a
height of 11 $\mu$m.

\subsection{New light extraction maps}

\begin{figure}[tbp]
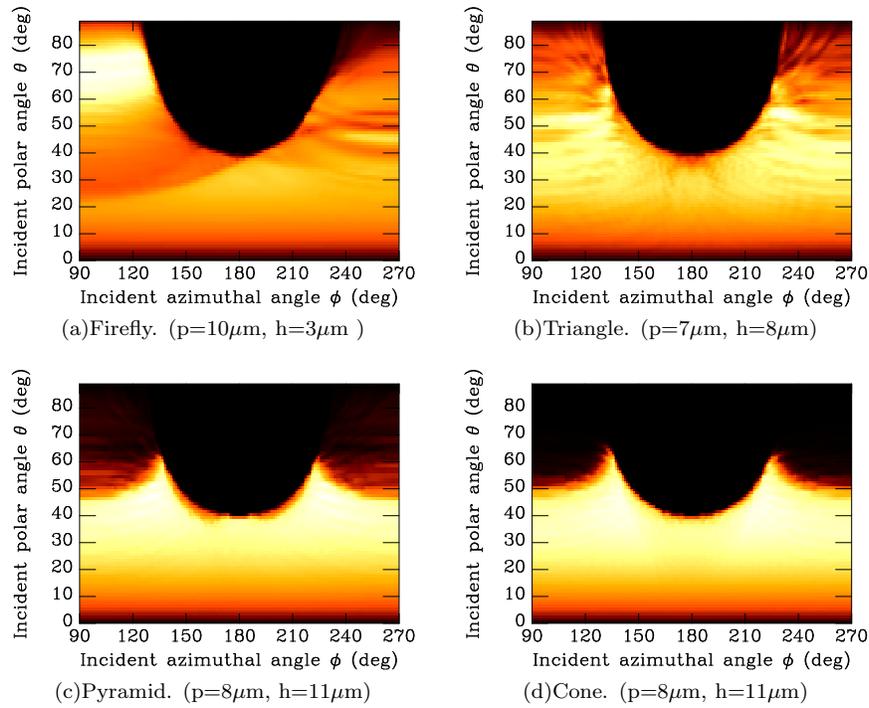

\centering \subfigure[Firefly. (p=10$\mu$m, h=3$\mu$m
)]{\includegraphics[height=4cm]{map_scales.eps}}\qquad
\subfigure[Triangle. (p=7$\mu$m, h=8$\mu$m)] {\includegraphics[height=4cm]{map_tri-2D.eps}}\\
\subfigure[Pyramid. (p=8$\mu$m,
h=11$\mu$m)]{\includegraphics[height=4cm]{map_tri-3D.eps}}\qquad
\subfigure[Cone. (p=8$\mu$m, h=11$\mu$m)]{\includegraphics[height=4cm]{map_d-sph.eps}}\\
\caption{Light extraction maps for the different analyzed
structures. For the two-dimensional structures, the light extraction is widely enhanced above the critical angle. The three-dimensional structures are less effective. The color scale is relative: black
shows zero transmission, and white shows maximal transmission.} \label{maps}
\end{figure}

The light extraction maps show more differences. Fig. \ref{maps}(a)
shows the light extraction for the firefly, clearly exhibiting the
huge improvement on the asymmetric light extraction. Fig.
\ref{maps}(b) shows the light extraction map for the two-dimensional
triangle-shaped structure. This shows a symmetric light extraction
as expected from the symmetric shape. Fig. \ref{maps}(c) shows the
light extraction map for the pyramid structure. One can clearly see
that the light extraction is not as efficient as in the firefly or
the triangle cases. The light is able to escape a little bit above
the critical angle, but not in such a spectacular way as in the
previous cases. Cones have a similar behavior as the pyramid
structure.

\subsection{Summarizing the highest extraction efficiencies}

Let's now compare the extraction efficiency for the most effective
geometry for each new structure (except for the firefly, where we
will stay with the geometry close to nature). For our appreciation,
we will choose the most effective direction of the azimuthal angle.
For the firefly structure, we will keep the periodicity of 10 $\mu$m
and the height of 3 $\mu$m. The triangle shows the best extraction
for a period of 7 $\mu$m and a height of 8 $\mu$m. These two
structures are most effective for an incident azimuthal angle of
$\phi=90^\circ$. The pyramid structure shows best results with a
periodicity of 8 $\mu$m and a height of 11 $\mu$m for an incident
azimuthal angle of $\phi=125^\circ$. The conical structure has been
shown to provide extraction very close to that of the pyramid
structure and is therefore not shown. Clearly, the firefly achieves
the best extraction behind the critical angle, but is a little bit
less effective for smaller incidences. The triangle structure is
evenly distributed over the whole range of incident polar angle, but
doesn't reach an extraction percentage as high as the firefly. The
pyramid structure shows the best extraction efficiency around the
critical angle, but falls quickly to nearly 0\% at larger
incidences. The integrated values of these curves gives 34.2\%,
26.6\% and 24.5\% for the firefly, the triangle and the pyramid
respectively.

\begin{figure}[h]
\begin{center}
\includegraphics[height=7cm]{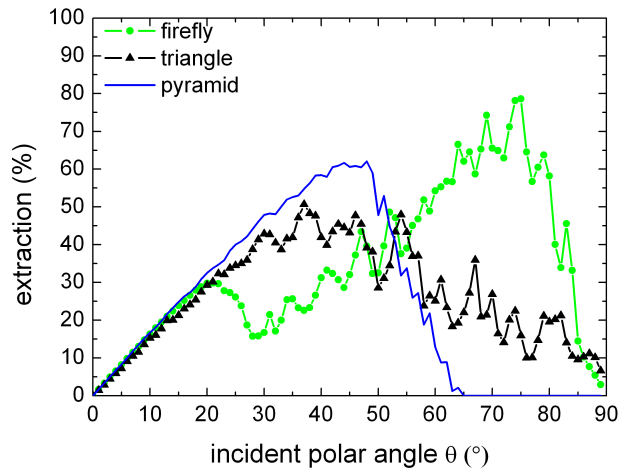}
\end{center}
\caption{Comparison between the light extraction in the best
geometry and the most effective direction for the different
structures. The firefly extracts the light in the most efficient way.}\label{comparison}
\end{figure}

%--------------------------------------------------------------------------------------
%       Conclusion
%--------------------------------------------------------------------------------------
\section{Conclusion}

The two-dimensional triangle-shaped structure provides extraction
efficiencies quite close to those exhibited by the firefly
structure. But the firefly structure still remains the best
extractor. The present study suggest that two-dimensional profiles
such as firefly or triangle structures provide light extraction
with a mechanism very different from three-dimensional such as
pyramids or cones. Surprisingly the studied three-dimensional
structures are not more efficient than two-dimensional structures. A
possible explanation is that the three-dimensional sharp points in
pyramids or cones appear with a surface density which is extremely
weak.

\acknowledgments

We thank Dr. Jean-Fran\c{c}ois Colomer for his help in the
scanning-electron microscopes operation. J.P.V. thanks Dr. Alan R.
Gillogly, Curator of Entomology at the Orma J. Smith Natural History
Museum, Caldwell, Idaho, USA, and Dr. Donald Windsor for for help in
collecting the specimens of fireflies used in the present study.
Both authors acknowledge very helpful discussions with Prof. Helen
Ghiradella (State University of New York, Albany, USA), Dr. Laure
Bonnaud (Museum National d'Histoire Naturelle, Paris France), Prof.
Laszl\`{o} Bir\`{o} (Nanostructures Department MFA, Research
Institute for Technical Physics and Materials Science MTA, Hungarian
Academy of Sciences, Budapest), Dr. Pol Limbourg (Royal Belgian
Institute of Natural Sciences, Brussels, Belgium), Dr. Gary Hevel
and Dr. Warren Steiner (National Museum of Natural History,
Smithsonian Institution, Washington D.C., USA). The project was
partly funded by the Action de Recherche Concert\'{e}e (ARC) Grant
No. 10/15-033 from the French Community of Belgium. The authors also
acknowledge using resources from the Interuniversity Scientific
Computing Facility located at the University of Namur, Belgium,
which is supported by the F.R.S.-FNRS under convention No.
2.4617.07. A.B. was supported as a Ph.D. student by the Belgian Fund
for Industrial and Agricultural Research (FRIA).

\bibliography{SPIE}   %>>>> bibliography data in spie.bib

\begin{thebibliography}{10}

\bibitem{Parker-book-2003}
Parker, A.,  [{\em In the blink of an eye: how vision kick-started the big bang
  of evolution}{\nolinebreak\hspace{0.1em}]}, The Free Press, Simon {\&}
  Schuster UK Ltd., London, UK, 1th~ed. ((2003)).

\bibitem{Vico-jpp-2008}
Vico, L.~D., Liu, Y.-J., and Lindh, R., ``Ab initio investigation on the
  chemical origin of the firefly bioluminescence,'' {\em Journal of
  Photochemistry and Photobiology A: Chemistry}~{\bf 194},  261--267 (2008).

\bibitem{Alam-NL-2012}
Alam, R., Fontaine, D.~M., Branchini, B.~R., and Maye, M.~M., ``Designing
  quantum rods for optimized energy trasfer with firefly luciferase enzymes,''
  {\em Nano Lett.}~{\bf 12},  3251--3256 (2012).

\bibitem{Ghiradella-I-1998}
Ghiradella, H., ``The anatomy of light production: The fine structure of the
  firefly lantern,'' {\em Microscopic Anatomy of Invertebrates}~{\bf 11A:
  Insecta},  363--381 (1998).

\bibitem{Hanna-JUR-1976}
Hanna, C.~H., Hopkins, T.~A., and Buck, J.~B., ``Peroxisomes of the firefy
  lantern,'' {\em J. Ultrastruct. Res.}~{\bf 57},  150--162 (1976).

\bibitem{Locke-JCB-1971}
Locke, M. and McMahon, J.~T., ``The origin and fate of microbodies in the fat
  body of an insect,'' {\em J. Cell Biol.}~{\bf 48},  61--78 (1971).

\bibitem{Smalley-JHC-1980}
Smalley, K.~N., Tarwater, D.~E., and Davidson, T.~L., ``Localization of
  fluorescent compounds in the firefly light organ,'' {\em J. Histochem.
  Cytochem.}~{\bf 28},  323--359 (1963).

\bibitem{Bay-SPIE-2009}
Bay, A. and Vigneron, J.~P., ``Light extraction from the bioluminescent organ
  of fireflies,'' {\em Proceedings of SPIE conference 7401: Biomimetics and
  Bioinspiration} ,  64860H (2009).

\bibitem{Vigneron-SPIE-2006}
Vigneron, J.~P. and Lousse, V., ``Variation of a photonic crystal color with
  the \textsc{M}iller indices of the exposed surface,'' {\em Proc. SPIE}~{\bf
  6128},  61281G (2006).

\bibitem{Hastings-Gene-1996}
Hastings, J.~W., ``Chemistries and colors of bioluminescent reactions: a
  review,'' {\em Gene}~{\bf 173},  5--11 (1996).

\bibitem{Vigneron-PRE-2005}
Vigneron, J.~P., Colomer, J.-F., Vigneron, N., and Lousse, V., ``Natural
  layer-by-layer photonic structure in the squamae of \textit{Hoplia coerulea}
  (\textsc{C}oleoptera),'' {\em Phys. Rev. E}~{\bf 72},  653--654 (2005).

\bibitem{Parker-nat-2003}
Parker, A.~R., Welch, V.~L., Driver, D., and Martini, N., ``Structural colour:
  Opal analogue discovered in a weevil,'' {\em Nature}~{\bf 426},  786--787
  (2003).

\bibitem{Vukusic-nat-2000}
Vukusic, P., Sambles, J.~R., and Lawrence, C.~R., ``Colour mixing in wing
  scales of a butterfly,'' {\em Nature}~{\bf 404},  457 (2000).

\bibitem{Zi-PRE-2006}
Zi, J., Yin, H., Shi, L., Sha, J., Li, Y., Qin, Y., Dong, B., Meyer, S., Liu,
  X., and Zhao, L., ``Iridescence in the neck feathers of domestic pigeons,''
  {\em Phys. Rev. E}~{\bf 74},  051916 1--6 (2006).

\bibitem{Jin-OE-2012}
Jin, Y., Yang, F., Li, Q., Zhu, Z., Zhu, J., and Fan, S., ``Enhanced light
  extraction from a \textsc{G}a\textsc{N}-based green light emitting diode with
  a hemicylindrical linear grating structure,'' {\em Opt. Exp.}~{\bf 20}(14),
  15818 (2012).

\bibitem{Lai-DRM-2011}
Lai, F.-I., Hsieh, Y.-L., and Lin, W.-T., ``Enhancement in the extraction
  efficiency and resisting electronic discharge ability of
  \textsc{G}a\textsc{N}-based light emitting diode by naturally grown textured
  surface,'' {\em Diamond \& related materials}~{\bf 20},  770--773 (2011).

\bibitem{Uthir.-MSEB-2010}
Uthirakumar, P., Kang, J.~H., Ryu, B.~D., Kim, H.~G., Kim, H.~K., and Hong,
  C.-H., ``Nanoscale \textsc{ITO}/\textsc{Z}n\textsc{O} layer-texturing for
  high-efficiency \textsc{I}n\textsc{G}a\textsc{N}/\textsc{G}a\textsc{N} light
  emitting diodes,'' {\em Material Science and Engineering B}~{\bf 166},
  230--234 (2010).

\end{thebibliography}
\bibliographystyle{spiebib}   %>>>> makes bibtex use spiebib.bst

\end{document}